\documentclass{ws-procs961x669}
\usepackage[mathscr]{euscript}
\usepackage{array,multirow,makecell} 
\usepackage[breaklinks, colorlinks, citecolor=blue]{hyperref}
\usepackage{url,doi}
\usepackage{graphicx}
\usepackage{epstopdf}
\usepackage{bm}
\usepackage{booktabs}

\begin{document}


\title{Structure evolution with cosmic backgrounds from radio to far infrared\footnote{Based on talks presented at the Seventeenth Marcel 
            Grossmann Meeting on General Relativity, July 2024.}}

\author{$\;\;\;\;\;\;\;\;\;\;\;\;$Carlo Burigana;$^{1,2,a}$ 
Tiziana Trombetti;$^{1,b}$
Matteo Bonato;$^{3,c}$\newline
Ad{\'e}lie Gorce;$^{4,d}$
Luigi Toffolatti$^{5,6,e}$
}
\address{$^{1}$INAF--IRA, Via Piero Gobetti 101, 40129 Bologna, Italy
\footnote{Istituto Nazionale di Astrofisica --
Istituto di Radioastronomia, Via Piero Gobetti 101, 40129 Bologna, Italy}\\
$^{2}$INFN, Sezione di Bologna, Via Irnerio 46, 40126, Bologna, Italy\\
$^{3}$INAF--IRA and Italian ALMA Regional Centre, Via Piero Gobetti 101, 40129 Bologna, Italy\\
$^{4}$Institut d'Astrophysique Spatiale, B\^atiment 121 - Universit\'e Paris-Saclay,\\Rue Jean-Dominique Cassini, 91405 Orsay cedex, France\\
$^{5}$Departamento de F\'{i}sica Universidad de Oviedo, C. Federico Garc\'{i}a Lorca 18,\\33007 Oviedo, Spain\\
$^{6}$Instituto Universitario de Ciencias y Tecnolog\'{i}as Espaciales de Asturias (ICTEA),\\C. Independencia 13, 33004 Oviedo, Spain \\
\vskip 0.1cm
$^{a}$carlo.burigana@inaf.it $-$ $^{b}$tiziana.trombetti@inaf.it $-$ $^{c}$matteo.bonato@inaf.it\\
$^{d}$adelie.gorce@universite-paris-saclay.fr $-$ $^{e}$ltoffolatti@uniovi.es}

\begin{abstract}
Cosmic background radiation, both diffuse and discrete in nature, produced at different cosmic epochs before and after recombination, provides key information on the evolution of cosmic structures. 
We discuss the main classes of sources that contribute to the extragalactic background light from radio to sub-millimetre wavelenghs and the currently open question on the level of the cosmic radio background spectrum. The redshifted 21cm line signal from cosmological neutral Hydrogen during the primeval phases of cosmic structures as a probe of the cosmological reionisation process is presented, along with the route for confident detection of this signal. We then describe the basic formalism and the feasibility to study via a differential approach, based mainly on dipole analysis, the tiny imprints in the CB spectrum expected from a variety of cosmological and astrophysical processes at work during the early phases of cosmic perturbation and structure evolution. Finally, we discuss the identification of high-redshift sub-millimetre lensed galaxies with extreme magnifications in the {\it Planck} maps and their use for the comprehension of fundamental processes in early galaxy formation and evolution.
\end{abstract}


\bodymatter

\def\dn{{(n)}}
\def\lsim{\,\lower2truept\hbox{${<\atop\hbox{\raise4truept\hbox{$\sim$}}}$}\,}
\def\gsim{\,\lower2truept\hbox{${> \atop\hbox{\raise4truept\hbox{$\sim$}}}$}\,}

\section{Introduction}
\label{sect:intro}

The background radiation of cosmic or extragalactic origin (CB) is all the radiation incident on Earth's atmosphere and coming from outside the Milky Way as well as from outside the Local Group of galaxies.
As is well known, and also very precisely measured, the Cosmic Microwave Background (CMB), the nearly blackbody (BB) radiation left over from the very hot early phase of the Universe, is, by far, the dominant part of the CB at centimetre (cm) to millimetre (mm) wavelengths. The most remarkable last scattering of CMB photons by matter occurred at about 380,000 years after the Big Bang,\cite{Planck2018} when the temperature cooled enough to allow protons to combine with electrons and form neutral atoms, during what is commonly called as the recombination, transforming the Universe from optically thick to optically thin at a redshift, $z$, of 
$\simeq 1100$. On the other hand, a further scattering of CMB photons by matter occurred at later epochs, during the so-called Epoch of Reionisation (EoR) when matter became ionised again because of radiation and energy sources associate to the emerging of first stars and galaxies.

The CMB shows a peak of intensity, $I_\nu$, at a frequency $\nu$ of about 220\,GHz, or at a wavelength $\lambda$ of 1.4\,mm, in terms $\nu I_\nu$ and a best fit current effective temperature 
$T_0 = (2.72548 \pm 0.00057)$\,K\cite{2009ApJ...707..916F} in the BB spectrum approximation such that $aT_0^4$, with $a = 8\pi I_3 k^4 / (hc)^3$, $I_3 = \pi^4/15$, gives the current CMB energy density,
$k$, $h$ and $c$ being the Boltzmann and Planck constants and the speed of light. The accurate mapping of CMB anisotropy in both total intensity (or temperature) and polarization (in $Q$ and $U$ linear polarization Stokes parameters, and, in the future, possibly also in the circular polarization $V$) already provided a wealth of information about the properties of the Universe and on the formation and evolution of its structures, also driving the development and implementation of improved data analysis approaches aimed at refining the treatment of calibration and systematic effects and at separating the cosmological signal from the astrophysical foreground emissions through a global analysis of the sky signals from radio to infrared (IR), in order to derive more faithful constraints of cosmological models and parameters, as discussed by I.~K. Wehus in this meeting. 

The accurate investigation of the currently poorly known tiny deviations of the CMB monopole spectrum from a BB, and, in general, of the details of the CB monopole frequency spectrum represents a new window for cosmological investigations.
They are usually observationally studied through direct absolute measurements. Owing precise relative and inter-frequency calibrations, the CB monopole spectrum can be also studied via differential measurements looking at the frequency dependence of the pattern at higher multipoles, particularly of the dipole, induced by the observer motion with respect to a reference frame at rest with respect to the considered CB. The observed extragalactic background radiation has been produced at different cosmic epochs after the recombination and, thus, it can give direct information on the evolution with cosmic time of the underlying sources, both diffuse and discrete in nature. 

We focus here on four main topics.
In Sect. \ref{sect:ES}, we discuss the extragalactic source background and how the main classes of sources contribute to the extragalactic background light (EBL), together with the controversial question of the level of 
cosmic radio background (CRB) spectrum.
Sect. \ref{sect:EoR} concerns the study of the cosmological reionisation process through the redshifted 21cm line signal from cosmological neutral Hydrogen (HI) produced during the primeval phases of cosmic structures, discussing the complexity of this high-$z$ signal detection and the different avenues to overcome it.
In Sect. \ref{sect:spectrum}, we deal with the formalism to predict how the intrinsic CB monopole frequency spectrum is modified and transferred to higher multipoles in the presence of an observer motion and the possibility to constrain (or even detect) via differential approaches the tiny imprints in the CB spectrum expected from a variety of cosmological and astrophysical processes at work during the early phases of cosmic perturbation and structure evolution. Finally, in Sect. \ref{sect:lens}, we discuss how the {\it Planck} mission all-sky survey has proven useful for identifying rare high-$z$ sub-mm lensed galaxies with extreme magnifications that are crucial for the comprehension of fundamental processes in early galaxy formation and evolution.

\section{Extragalactic source background}
\label{sect:ES}

At radio to sub-mm wavelengths the EBL is mainly contributed by photons produced by various astrophysical processes inside discrete or compact extragalactic sources, 
that, except for facilities with extremely high resolution, are typically observed as point sources inside the telescope beam.
Excluding the spectral region where the CMB dominates, Active Galactic Nuclei (AGN) and star-forming galaxies (SFG)\footnote{See Sect. \ref{sect:lens} for perspectives based on strong lensing.} produce the most relevant contribution to the CB.

Direct detailed measurements of the EBL at many frequencies can be compared with the total background radiation indirectly derived by integrating the differential number counts of sources, $N(S_\nu)$, 
down to the faintest flux density limits, $S_{\rm min}$, $\int^{S_{\rm max}}_{S_{\rm min}} S_\nu N(S_\nu) d\nu$.
Direct measurements integrates the total sky light inside the telescope beam and are followed by the subtraction of foregrounds, such as the Zodiacal Light and diffuse Galactic light (in the far-IR and sub-mm), and also by the subtraction of radiation emerging from the Galactic halo and, finally, by the removal of the CMB. 
Indirect estimates include the measurement and modelling of source number counts, that are only sensitive to discrete sources of radiation (i.e., normal, SFG and AGN) but have the advantage of not being severely contaminated by foregrounds. In principle, the comparison between the results obtained with the two methods could allow us to constrain the history of star formation, the evolution of supermassive black holes and also the role of dust processing, as well as to grasp the possible presence of unknown classes of sources or diffuse emissions.

The analysis of the CB at sub-mm to IR wavelengths, the so-called Cosmic Infrared Background (CIB), presented in this meeting by A. Maniyar, is crucial for the understanding of the star formation history. The CIB spectral shape and fluctuations together with the strong correlation between galaxy stellar mass and obscured star formation suggest that the CIB is largely contributed by known classes of galaxies in which a large fraction of the optical and ultraviolet light from stars is reprocessed at longer wavelengths by dust.
At mm wavelengths, a recent analysis of Atacama Large Millimeter/submillimeter Array (ALMA) data in the 
Hubble Space Telescope (HST) Ultra Deep Field (HUDF) indicates that the contribution from detected sources plus with that derived from a stacking analysis of a shallower map on the positions of James Webb Space Telescope (JWST) galaxies undetected by ALMA is consistent, within uncertainty, with the background: the contribution from faint undetected objects is found to be small and converging, suggesting that JWST has detected essentially all galaxies contributing to the CIB.\cite{2024MNRAS.528.5019H}

At radio wavelengths, upcoming surveys that will be carried out in the next future with the new facilities and, especially, with the Square Kilometre Array\footnote{SKA is a giant interferometer consisting of a low-frequency array in Western Australia and another mid-frequency array in the South-African Karoo desert.} (SKA), will likely be able to make number counts convergent at least at all radio frequencies below $10\div20$\,GHz. On the other hand, it is currently still necessary to use specific cosmological evolution models for extragalactic sources to extrapolate down in flux their contribution to the CB radiation, since the available surveys are not able to reach sufficiently low flux density limits for the cosmic radio background (CRB) (from source counts data) to be convergent, as recently discussed in Ref. \citenum{2023MNRAS.521..332T}. 

\subsection{Supermassive black hole evolution and radio-loud AGN}
 \label{sect:evolRS}
 
Extragalactic sources that prevail in the microwave sky are fuelled by AGN. Their observed flux density comes from synchrotron radiation due to accelerated relativistic charged particles and shows a frequency spectrum that is typically characterized by a power law with $S\propto\nu^{\alpha}$: emission emerging in extended radio lobes exhibits a steep spectrum (with slope $\alpha<-0.5$) while that from compact regions of radio jets 
shows a flat spectrum ($-0.5<\alpha<0.5$).
In the \lq AGN unified model\rq\,\cite{urr95,net15} the main difference between the two populations is due to the different orientation of the observer line of sight and of the axis of the jet emerging from the central black hole, a side-on view of the jet-axis resulting into a steep spectrum source, a line of sight parallel to the jet-axis resulting into a flat spectrum compact source (i.e. a blazar).\citep{dez10} The steep spectrum population is most abundant at $\nu < 10\div20$ GHz but becomes sub-dominant at higher frequencies, more suitable for CMB anisotropy experiments, where the flat spectrum population dominates and represents, with its emission with a  polarisation degree of a few percent, the most important foreground contamination in CMB polarization experiments at sub-degree angular scales. 
In the flat spectrum source model of Ref.\,\citenum{tuc11}, updated in Ref.\,\citenum{lag20}, the spectrum is expected to break at some frequency, $\nu_{\rm M}$, between 10 and 1000 GHz and to steepen at higher frequencies, due to the transition of the observed synchrotron emission from the optically thick to the optically thin regime and to electrons cooling effects. In particular, this break frequency appears at $\nu \gsim100$ GHz in BL-Lacs and at $\nu < 100$ GHz in flat spectrum radio quasars (FSRQs), implying less compact emitting regions than in BL-Lacs. Indeed, the distance, $r_{\rm M}$, from the AGN core of the jet portion that dominates the emission at frequency $\nu_{\rm M}$ defines the dimension and, thus, the compactness of the emitting region at that frequency, and it is the most relevant parameter in the estimate of $\nu_{\rm M}$, being approximately $\nu_{\rm M} \propto r_{\rm M}^{-1}$ (as in predictions of inhomogeneous jet models). In the \lq C2Ex\rq\, model, the most successful model discussed in Ref. \citenum{tuc11}, it is assumed that $0.01 < r_{\rm M} / {\rm pc} < 0.3$ for BL-Lacs and a more extended emitting region, $0.3 < r_{\rm M} / {\rm pc} < 10$, for FSRQs. This model is able to give a good fit to the observational number counts of radio sources in the whole frequency range $30 \le \nu / {\rm GHz} \le 220$, in both flux density\cite{tuc11} and polarisation\cite{lag20} (also for blazars detected in Herschel Astrophysical Terahertz Large Area Survey (H-ATLAS), see Ref. \citenum{2022MNRAS.513.6013M}). 

The luminosity functions (LFs) of radio-loud AGN is driven by the cosmological evolution of the supermassive black hole (SMBH) hosted in their nuclei. 
According to Ref. \citenum{2014ARA&A..52..589H}, there are two different accretion \lq modes\rq\, in AGN, the radiative-mode, in which the potential energy of gas is efficiently converted to radiation, and the jet-mode, which is radiatively inefficient and in which most of the energy output is in form of kinetic energy (two-sided jets). Depending on the Eddington ratio, $\lambda_{\rm E} = L_{\rm bolometric} / L_{\rm Eddington}$, black holes accrete in three distinct \lq modes\rq: at low $\lambda_{\rm E}$, only a radiatively inefficient, kinetically dominated mode is allowed [low kinetic (LK)]; at high $\lambda_{\rm E}$, AGN may display a purely radiative [radio quiet, high radiative (HR)] as well as a kinetic [radio loud, high kinetic (HK)] mode.\cite{2008MNRAS.388.1011M}

In this framework, relying on physical and phenomenological relations to statistically calculate the radio luminosity of AGN cores, corrected for beaming effects, by linking it with the SMBH at their centre, through the fundamental plane of black hole activity,\cite{2003MNRAS.345.1057M,2004A&A...414..895F} and computing radio luminosity from extended jets and lobes through a power-law relationship according to the source compactness, Ref. \citenum{2021A&A...650A.127T} developed a model for the LFs of the radio-loud AGN at GHz frequencies. The model consider two main classes, the LK \lq mode\rq\, for $\lambda_{\rm E} \le 0.01$ and the HK \lq mode\rq\, for $\lambda_{\rm E} \ge 0.01$,
and involves only few free parameters that are estimated by fitting two different observational data sets: local, or low-$z$, LFs of radio-loud AGN at 1.4 GHz and differential number counts of extragalactic radio sources at 1.4 and 5 GHz. Locally, the LK \lq mode\rq\, AGN is found to be dominant at low luminosities.
In spite of its simplicity and except for a certain underestimation -- but only at low radio luminosities -- of some recent measures of the HK mode AGN LF at low-$z$, the model well reproduces almost all published data on LFs of LK mode AGN and of the total AGN population at $z \le 1.5$ and also in the full range of luminosities currently probed by data: remarkably, this holds also regarding observational data not included in the fit to determine the free parameters.

\subsection{Cosmic radio background}
 \label{sect:back}
 
As is well known, the averaged temperature of the CB data at 1\,GHz\,$\lsim \nu \lsim$\,30\,GHz is slightly below the Cosmic Background Explorer/Far Infrared Absolute Spectrophotometer (COBE/FIRAS) temperature determination at $\nu \gsim$\,30\,GHz. On the other hand, the measurements below 1 GHz, see e.g. Ref. \citenum{2018ApJ...858L...9D}, and the excess\cite{2011ApJ...734....5F} at $\simeq 3.3$\,GHz claimed by the Absolute Radiometer for Cosmology, Astrophysics, and Diffuse Emission\cite{2011ApJ...730..138S} (ARCADE 2) indicate a remarkable temperature increase in the radio tail of the CB radiation (see e.g. Appendix A in Ref. \citenum{2020PhRvR...2a3210B} for a recent data compilation). The amplitude and steepness of the CRB seems to exclude an interpretation of the excess driven by a substantial cosmological diffuse free-free (FF) emission (see also Sect. \ref{sect:spectrumobs}), leaving room for possible contributions\cite{2011ApJ...734....6S} coming from an underestimated Galactic foreground, an unknown class of distant and faint radio sources,\footnote{This scenario was also considered to explain the claimed detection of a very deep absorption profile in the redshifted 21cm line signal (see Sect. \ref{sect:EoR}).} or a truly cosmological signal.

Recently, Ref. \citenum{2023MNRAS.521..332T} have revisited the current constraints on the CRB at cm and mm wavelengths from source number counts across a broad frequency range, relying on earlier studies and on the largest available databases (see also Refs. \citenum{2003NewAR..47..357W} and \citenum{dez10}), and modeled the EBL at radio frequencies by considering both AGN\footnote{See Sect. 3.1 in Ref.  \citenum{2024A&A...684A..82T} for a discussion about the consistency of the EBL at GHz frequencies estimated in Refs. \citenum{tuc11} and \citenum{2023MNRAS.521..332T}.} and SFG populations.
The two populations are found to contribute almost equally to the CRB energy and, in particular, only at 3 GHz the source count contributions to the EBL is found to converge, as indicated by a $P(D)$ fluctuation analysis,
\cite{2014MNRAS.440.2791V} while empirical lower limits to the radio EBL are provided in the other bands. The Shark\cite{2018MNRAS.481.3573L} semi-analytic model for SFG population and the ProSpect\cite{2020MNRAS.495..905R} spectral energy distribution (SED) code to characterise the ultraviolet-optical-infrared-mm-radio SFG EBL at all frequencies from the cosmic star-formation history and the assumption of a Chabrier initial mass function\cite{2003PASP..115..763C} have been adopted in Ref. \citenum{2023MNRAS.521..332T}. Remarkably, there appears to be a currently unsolvable discrepancy of a factor $3 \div 5$ between the (high) direct measurements of CRB and the (low) integrated background signal from the most recently published source count analyses: a significant missing discrete source population is ruled out\footnote{As discussed in Ref. \citenum{2023MNRAS.521..332T}, with current models of source number counts -- or even reaching fainter flux densities, down to $\simeq (10 \div 100)$\,nJy, by current as well
as future facilities (e.g. ALMA and SKA arrays) -- the possibility that the whole of the measured ARCADE 2 temperature could come from galaxies and AGN alone is increasingly unlikely.} and, thus, the cause of this discrepancy can be due to the foreground subtraction or in a yet unknown genuine diffuse component,\cite{2023MNRAS.521..332T} possibly of cosmological origin, as attempted e.g. in the perspective of non-equilibrium statistical mechanics.\cite{2020PhRvR...2a3210B}

\section{The Epoch of Reionisation through the low-frequency radio lens}
\label{sect:EoR}

\begin{figure}[t!]
\center
\includegraphics[width=8cm]{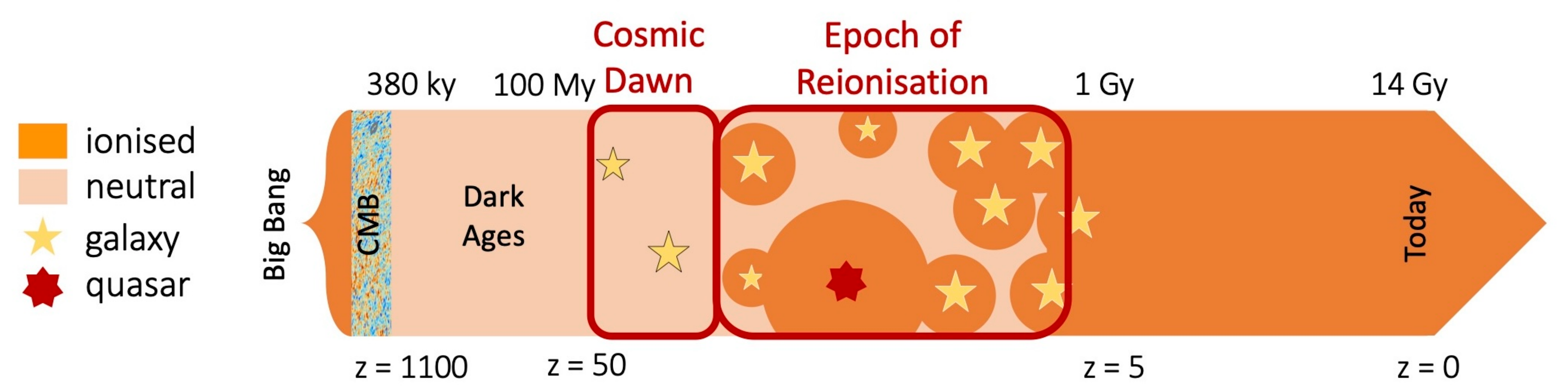}
\includegraphics[width=11cm]{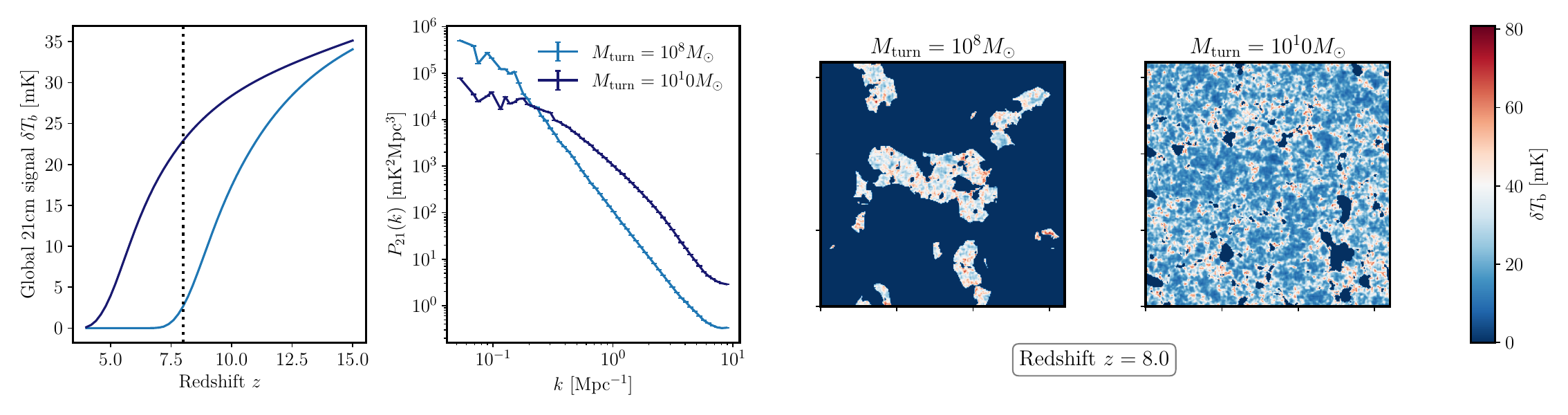}
\caption{{\it Top}: simplified timeline of the large-scale evolution of the Universe. {\it Bottom}: the 21cm global signal, power spectrum, and its intensity map at $z=8$.
The two models correspond to different values of the minimal mass of dark matter haloes required from them to form stars.\cite{MesingerFurlanetto_2011}}
\label{fig:21cm}
\end{figure}

After the recombination of primordial atoms, 
during what is commonly referred to as the Dark Ages, the Universe is mostly neutral. As the first galaxies form, their radiation progressively ionises the Hydrogen atoms in the surrounding intergalactic medium (IGM) in a period called the EoR. Each source produces an ionised bubble which grows until all bubbles percolate and the
IGM becomes fully ionised.
Studying the chronology and topology of the EoR has the potential to tell us much about the nature of the early Universe, of its first stars, quasars, and galaxies. Indeed, as illustrated in Fig.~\ref{fig:21cm} (top panel), depending on the nature of the source, the shape and size of the ionised bubble it produces varies. Reciprocally, a denser or `clumpier' IGM will slow down the ionisation front, resulting in long-lasting pockets of HI.

Looking at the HI absorption troughs in quasar spectra, estimating the ionising photon budget with deep galaxy surveys, or measuring the amount of CMB photons scattered by the free electrons produced during reionisation through its optical depth, can already give some loose constraints on the timeline of the EoR, that is, on the $z$-evolution of the IGM ionised fraction. From the former, we know reionisation might have ended by $z= 5.3$.\cite{BosmanDavies_2022} From the latter, we know $50\%$ of the IGM might be ionised by $z = 7.67 \pm 0.73$.\cite{Planck2018} However, both results rely on strong model assumptions and give an incomplete view of the process.

Observations of the high-$z$ 21cm signal will lift these uncertainties and enable a direct measurement of the IGM ionised fraction, whether averaged over the whole sky, or locally.
This signal corresponds to the emission of a photon with a wavelength of exactly 21cm during the hyperfine transition of the HI atom.
This is a forbidden transition, but the large amounts of HI in the IGM make the cosmological signal detectable. The redshifted 21cm line signal is observed in contrast to a background radiation, often taken to be the CMB, and its brightness temperature writes\cite{pritchard_loeb_2012}
\begin{equation}
\label{eq:dTb}
\delta T_{\mathrm{b}}(z)\simeq T_0(z)\, x_{\mathrm{HI}}\left(1+\delta_{\mathrm{b}}\right) \left[ 1 - \frac{T_\mathrm{CMB}}{T_\mathrm{S}} \right],
\end{equation}
where $x_\mathrm{HI}$ is the neutral fraction of the IGM, $\delta_b$ is the baryon overdensity, $T_S$ is the spin temperature, and we simply assume $T_0$ to be only dependent on the redshift and cosmological parameters. Hence, the intensity of the 21cm signal is a direct tracer of the baryon distribution and of the IGM ionisation. Through redshift, the 21cm photons emitted during the EoR are observed today at low radio frequencies and can be used to map the Universe at \emph{any} time.
Observing this signal at different scales provides a variety of information about reionisation, see Fig.~\ref{fig:21cm} (bottom panel).

The global 21cm signal is measured with a simple dipole. The first claimed measurement of this signal at $z\sim 17$ has published in 2018 
on the basis of the data from the Experiment to Detect the Global Epoch of Reionization Signature (EDGES).\cite{EDGES2018} However, this measurement differs from expectations\cite{pritchard_loeb_2012} in both amplitude and shape. Various explanations have been proposed to explain this discrepancy, from instrumental systematics\cite{SinghSubrahmanyan_2019} to extreme physical models.\cite{BarkanaOutmezguine_2018,Ewall-WiceChang_2018,MirochaFurlanetto_2019} However, the Shaped Antenna measurement of the background Radio Spectrum (SARAS) observations reject the EDGES best-fitting profile with $95.3\%$ confidence.\cite{SinghJishnu_2022} An independent measurement is necessary to confirm or infirm the EDGES claim.

The high-$z$ 21cm power spectrum (PS) carries information about the spatial fluctuations of the signal, giving access to not only a more precise measurement of the reionisation timeline, but also to constraints on the physical properties of early galaxies such as their ionising emissivity. Radio-interferometers around the world are targeting this measurement. The visibility measured by the baseline $\mathbf{b}_{i j} $ between antennas $i$ and $j$ of an interferometer at frequency $\nu$ writes
\begin{equation}
V_{i j}(\nu)=\int B_{i j}(\hat{\mathbf{r}}, \nu) I(\hat{\mathbf{r}}, \nu) \exp \left[-2 \pi i \frac{\nu}{c} \mathbf{b}_{i j} \cdot \hat{\mathbf{r}}\right] d \Omega,
\label{eq:vis}
\end{equation}
where $\Omega$ is the solid angle, $\hat{\mathbf{r}}$ is the radial norm vector in the sky plane, $B_{ij}$ is the beam, and $I$ the signal intensity. Eq. \eqref{eq:vis} shows that the baseline $\mathbf{b}_{i j} $ is analogous to the Fourier dual of the sky angle, $\bm{k}_\perp$. When measuring the 21cm signal intensity, a (simple) estimator of its PS can conveniently be written as $\widehat{P}(\boldsymbol{k}) \propto\langle|\widetilde{V}_{ij}(\nu)|^{2}\rangle$.

Despite many dedicated telescopes around the world, the high-$z$ 21cm PS remains elusive: the current lowest upper limits, obtained by the Hydrogen Epoch of Reionization Array (HERA),\cite{HERACollaborationAbdurashidova_2023} are about 100 times larger than fiducial theoretical models,\cite{MesingerFurlanetto_2011} but sufficiently low to exclude extreme physical models of reionisation: HERA data implies that IGM was heated by $z = 10.4$, likely by high-mass X-ray binaries.\cite{HERACollaborationAbdurashidova_2023} 
Because of the large theoretical uncertainty, four independent models have been used to interpret the HERA measurements -- and they all concur.

SKA has ambitions to draw 21cm signal intensity maps thanks to its low-frequency array covering key frequencies for EoR studies. These maps, see Fig.~\ref{fig:21cm}, will give access to detailed information about the morphology of ionised bubbles and, hence, about the physical properties of high-$z$ galaxies.
They will complement PS-based analyses: the 21cm signal is expected to exhibit strong non-Gaussianities, which are, by definition, not catched by the PS.
To extract a maximum of information from such maps whilst taking into account specific limitations of the data, efficient tools have been recently developed, including Minkowski functionals \& topological functions,\cite{ElbersvandeWeygaert_2019} higher order statistics \& the bispectrum,\cite{GorcePritchard_2019} machine-learning-based techniques,\cite{NeutschHeneka_2022} scattering transforms,\cite{HothiAllys_2024} and one-point statistics.\cite{GorceHutter_2021}

\subsection{When will we observe the high-redshift 21cm signal?}
\label{sec:when}

Despite the potential of the high-$z$ 21cm signal to understand the EoR and the large-scale history of the Universe, no measurement has yet been confirmed. Going from a low-frequency radio observation to an estimate of the cosmological 21cm signal is indeed very difficult. Here, we list the major issues, with a focus on HERA.

The frequencies targeted by most 21cm experiments range from $\sim 50$\,MHz to 250\,MHz, corresponding to $5 < z < 30$. 
Most of the target frequency band is polluted by human emissions  -- or radio frequency interference (RFI), such as aviation communication, FM radio ($12 < z< 15$), and radars. The RFI signal is so loud that the cosmological signal can hardly be recovered.
Therefore, most 21cm experiments are located in extremely remote, radio-quiet areas, from the South African Karoo desert\cite{HERACollaborationAbdurashidova_2023} to the Canadian High Arctic.\cite{ChiangDyson_2020}

Because of the large numbers of antennas making up the array, interferometers gather huge amounts of data. For example, one observing season of HERA corresponds to 160 8-hour long nights. Data is taken every 10.7\,s over 1536 frequency channels and along two directions corresponding to different polarisation. The signals from 350 antennas are correlated to form 122\,150 visibilities, such that each observing season corresponds to about $10^{12}$ measurements, or 170\,TB of data. 
The storage and analysis of such large data sets require sufficient computational power and advanced, custom real-time signal processing methods.\cite{LaPlanteWilliams_2021}

A key issue in the observation of the high-$z$ 21cm signal is the presence of extremely bright foregrounds, a hundred to a thousand times brighter than the cosmological signal.
These foregrounds consist of Galactic emissions such as free-free (FF) and synchrotron, as well as extragalactic radio point sources (see Sect. \ref{sect:ES}). Different methods have been developed to mitigate their impact, all relying on the assumption that foregrounds are spectrally smooth whilst the cosmological signal is not. One can either directly remove foreground signal from the observations,\cite{MertensGhosh_2018} which requires a good knowledge of these foregrounds to subtract their modelled contribution, or one can simply avoid them and target regions of Fourier space where their contribution is expected to be small.\cite{LiuParsons_2014} Although conceptually simple, these techniques are difficult to apply in practice, mostly because of instrumental systematics: the chromaticity of the beam introduces spectral structure in the initially smooth foreground signal.\cite{GorceGanjam_2023}
Another example is the high variability of the ionosphere and the difficulty to model its signal as a function of time and frequency.

The largest number of headaches in the 21cm analysis teams is caused by instrumental systematics that are largely unknown before observing starts but need to be accurately understood and characterised.
Otherwise, their unrefined removal from the data can lead to artificial upper limits and signal suppression. Examples are cross-coupling between antennas, signals measured in one antenna being reflected off another antenna and measured twice, and cable reflections.\cite{GanKoopmans_2022, RathPascua_2024}

Obtaining a confirmed observation is an extremely difficult task which can only be achieved incrementally.
If a first measurement of the global 21cm signal has been claimed in 2018,\cite{EDGES2018} it has not yet been confirmed. Meanwhile, as the upper limits on the 21cm PS get deeper, new systematics are uncovered, requiring months of work to be properly characterised.
Collaborating and sharing our experience between experiments around the world is the best course of action to reach detection (see Ref. \citenum{GorceThisproc} for further details).

\section{Modification and transfer of the cosmic background spectrum}
\label{sect:spectrum}

Boosting effects associated to the observer peculiar motion modify and transfer to higher multipoles the frequency spectrum of the background isotropic monopole, the largest effect, mainly attributed to the solar system motion, appearing in the dipole ($\ell = 1$).\cite{1981A&A....94L..33D} Their theoretical computation relies on the Compton-Getting effect,\cite{1970PSS1825F} based on the Lorentz invariance of the photon distribution function, $\eta(\nu)$. At the frequency $\nu$, the observed signal in equivalent thermodynamic temperature, $T_{\rm th} (\nu) ={(h\nu/k)} / {\ln(1+1/\eta(\nu))}$, is
\begin{align}\label{eq:eta_boost}
T_{\rm th}^{\rm BB/dist} (\nu, {\hat{n}}, \vec{\beta}) =
\frac{xT_{0}} {{\rm{ln}}(1 + 1 / (\eta(\nu, {\hat{n}}, \vec{\beta}))^{\rm BB/dist}) } = \frac{xT_{0}} {{\rm{ln}}(1 + 1 / \eta(\nu')) }\, ,
\end{align}
\noindent where 
$\eta(\nu, {\hat{n}}, \vec{\beta}) = \eta(\nu')$, with
$\nu' = \nu (1 - {\hat{n}} \cdot \vec{\beta})/(1 - \beta^2)^{1/2} \, $; here ${\hat{n}}$ is the sky direction unit vector, $\vec{\beta} = \vec{{\rm v}} / c$ the dimensionless observer velocity,  
$x=h\nu/(kT_r)$ the CMB redshift invariant dimensionless frequency, $T_r=T_0(1+z)$ the CMB redshift dependent effective temperature
and the notation `BB/dist' denotes a BB spectrum or any type of spectrum.\cite{2018JCAP...04..021B}

Eq.\,\eqref{eq:eta_boost} can be expanded in spherical harmonics, with coefficients that can be explicitly calculated up to the desired order of accuracy. Choosing a reference system with the $z$-axis parallel to the observer velocity, the dependence on the colatitude $\theta$ is kept while the one on the longitude $\phi$ vanishes, only the spherical coefficients $a_{\ell,m} (\nu, \beta)$ with $m=0$ do not vanish and the spherical harmonics reduce to the renormalized associated Legendre polynomials.
Considering the expansion up to a certain multipole, e.g. $\ell_{max} = 6$,\footnote{Since $\beta \simeq 1.2336 \times10^{-3}$,\cite{2020A&A...641A...1P} adopting $\ell_{\rm max}=6$ allows us to achieve a  numerical accuracy suitable for any application even in the very far future since the amplitude of $a_{\ell,m} (\nu, \beta)$ decreases at increasing multipole, $\ell$, as $\beta^{\ell \cdot p}$, with $p \approx 1$ (for a BB, $p=1$ and $a_{\ell,m} (\nu, \beta)$ does not depend on $\nu$).} and 
computing the signal through Eq.\,\eqref{eq:eta_boost} in only $N = \ell_{\rm max}+1$ sky directions defined by a set of different colatitudes $\theta_i$ with $i=0,N-1$
we can construct a linear system of $N$ equations in the $N$ unknowns $a_{\ell,0} (\nu,\beta)$, with $\ell=0,N-1$. 
Choosing the $N$ colatitudes $\theta_i$ symmetrically with respect to $\pi/2$ (for instance $w_i = {\rm cos} \, \theta_i = 1, \sqrt{2}/2, 1/2, 0, -1/2, -\sqrt{2}/2, -1$ to simplify the algebra), the symmetry of the renormalized associated Legendre polynomials with respect to $\pi/2$ allows to separate
the system in two subsystems, one for $\ell=0$ and even multipoles and the other for odd multipoles, improving the solution accuracy because neglecting higher $\ell > \ell_{\rm max}$ produce a mean error dominated only by $\ell_{\rm max}+2$ for even $\ell$ (or from $\ell_{\rm max}+1$ for odd $\ell$).\cite{2021A&A...646A..75T}
Ultimately, for any type of background spectrum, the solution for each $a_{\ell,0} (\nu,\beta)$ can be written in terms of a linear combination of sums and differences of the signals from Eq. \eqref{eq:eta_boost},
$T_{\rm th}^{\rm BB/dist} (\nu,\beta, w)$ (hereafter $T_{\rm th}^{\rm BB/dist} (w)$ for simplicity), at the adopted colatitudes\cite{2021A&A...646A..75T}
\begin{align}
\label{eq:struct_evenodd}
a_{\ell,0} & = A_\ell \sqrt{\frac{4\pi}{2\ell+1}} \Bigg[ d_{\ell,1} \left({T_{\rm th}^{\rm BB/dist} (w=1) \pm T_{\rm th}^{\rm BB/dist} (w=-1) }\right) \nonumber
\\ & + d_{\ell,2} \left({T_{\rm th}^{\rm BB/dist} (w=\sqrt{2}/2) \pm T_{\rm th}^{\rm BB/dist} (w=-\sqrt{2}/2) }\right) 
\\ & + d_{\ell,3} \left({T_{\rm th}^{\rm BB/dist} (w=1/2) \pm T_{\rm th}^{\rm BB/dist} (w=-1/2) }\right) + d_{\ell,4} T_{\rm th}^{\rm BB/dist} (w=0) \Bigg] \, \nonumber
\end{align}
\noindent where $\pm=+$ ($\pm=-$) for $\ell=0$ and even $\ell$ (for odd $\ell$) with $A_\ell$ and $d_{\ell,i}$ in Table \ref{tab:coeffsall}.

The values of $T_{\rm th}^{\rm BB/dist} (w)$ should be always very close to the one computed at $w=0$, i.e. perpendicularly to the observer motion.
Assuming that $T_{\rm th}^{\rm BB/dist} (w)$ can be expanded in Taylor's series around $w=0$, and denoting the derivatives of
$T_{\rm th}^{\rm BB/dist} (w)$ performed with respect to $w$ evaluated at $w=0$ with 
$T_{\rm th}^{(0)}$, $T_{\rm th}^{(1)}$, ..., $T_{\rm th}^{(6)}$, from order zero to order six, 
Eq. \eqref{eq:struct_evenodd}
becomes\cite{2024A&A...684A..82T} 
\begin{equation}
\label{eq:struct_der_ell}
a_{\ell,0} = {\frac{1}{D_\ell}} \sqrt{\frac{4\pi}{\ell+1}} \; \left[T_{\rm th}^{(\ell)}  + \frac{1}{s_{(\ell+2)}} T_{\rm th}^{(\ell+2)} + \frac{1}{s_{(\ell+4)}} T_{\rm th}^{(\ell+4)} + \frac{1}{s_{(\ell+6)}} T_{\rm th}^{(\ell+6)} \right] ,
\end{equation}
\begin{table}
\tbl{ \small For $\ell_{\rm max} = 6$, the table reports the $A_\ell$ and $d_{\ell,i}$ coefficients for the adopted colatitudes and the $s_{(n)}$ coefficients.}
{\begin{tabular}{@{}c|c|c|c|c|c|c|c|c@{}}
\toprule
      $\ell$ & $A_\ell$ & $d_{\ell,1}$ & $d_{\ell,2}$ & $d_{\ell,3}$ & $d_{\ell,4}$ & $s_{(\ell+2)}$ & $s_{(\ell+4)}$ & $s_{(\ell+6)}$ \\ [0.2ex]
        \hline
0 & 1/630 & 29 & 120 & 64 & 204 & 6 & 120 & 5040 \\[0.2ex]
1 & 1/210 & 29 & 60$\sqrt{2}$ & 32 & 0 & 10 & 280 & -- \\[0.2ex]
2 & 1/693 & 121 & 396 & $-$352 & $-$330 & 14 & 504 & --  \\[0.2ex]
3 & 2/135 & 13 & 15$\sqrt{2}$ & $-$56 & 0 & 18 & -- & -- \\[0.2ex]
4 & 8/385 & 9 & $-$10 & $-$16& 34 & 22 & -- & -- \\[0.2ex]
5 & 32/189 & 1 & $-$3$\sqrt{2}$ & 4 & 0 & -- & -- & -- \\[0.2ex]
6 & 64/693 & 1 & $-$6 & 8 & $-$6 & -- & -- & -- \\ [0.2ex]
    \bottomrule
    \end{tabular}
}
\label{tab:coeffsall}
\end{table}
\noindent
where $D_\ell = (2\ell-1) D_{\ell-1}$, $D_0 = 1$, and the coefficients $s_{(\ell+2)}$, $s_{(\ell+4)}$, $s_{(\ell+6)}$ are in Table \ref{tab:coeffsall}, the missing values reflecting the 
adopted $\ell_{\rm max}$ (or the consequent maximum derivative order). We note that: $(i)$ only the derivatives of even (odd) order contribute to $a_{\ell,0}$ for even (odd) $\ell$; $(ii)$ only the derivatives of order\,$\gsim \ell$ contribute to $a_{\ell,0}$; $(iii)$ the factors $1/s$ strongly decreases with the derivative order.\footnote{On the other hand, $(iii)$ and the typical overall scaling $a_{\ell,0} (\nu, \beta) \propto \beta^{\ell \cdot p}$ do not imply that at each multipole $\ell$ the terms from the derivatives of order greater than $\ell$ are in general not relevant, due to the different frequency dependencies of the derivatives of different orders.}

The relation between the $n$-th derivatives with respect to $w$ and $\nu'$ is
$\frac{{dT_{\rm th}^n}}{{dw^n}} =  \left[\frac{{-\beta \nu}}{ {(1 - \beta^2)^{1/2}}}\right ]^n\, \frac{{dT_{\rm th}^n}}{ {d\nu'^n}}$.
Specifying $w=0$ (or $\theta=\pi/2$)
${\frac{{dT_{\rm th}^n} }{ {d\nu'^n}}} \Bigm |_{w=0} = {\frac{{dT_{\rm th}^n} }{ {d\nu'^n}}} \Bigm|_{\nu'_{\beta,\perp}}$, where $\nu'$ is set to $\nu'_{\beta,\perp} = \nu /(1 - \beta^2)^{1/2}$.
For two different values of $\beta$ and $\beta_a$, the ratio of the above derivatives is
$\frac{{T_{\rm th}^\dn}|_{\beta}}{{T_{\rm th}^\dn}|_{\beta_a}} = f_a^{-n} \, \left[ { \frac{(1-\beta_a^2)} {(1-\beta^2)} } \right]^{n/2} \, R_n$,
with $f_a = \beta_a /\beta$, $R_n =  ( {{dT_{\rm th}^n} / {d\nu'^n}}|_{\nu'_{\beta,\perp}} ) \, ( {{dT_{\rm th}^n} / {d\nu'^n}}|_{\nu'_{\beta_a,\perp}} )^{-1}$, $\nu'_{\beta_a,\perp}$ defined for $\beta = \beta_a$.
Except for $T_{\rm th}$ strongly varying in frequency, only $f_a^{-n}$ can be remarkably different from unity.

The solutions discussed above allow the explicit and highly accurate calculation of the $a_{\ell,0} (\nu,\beta)$'s, and then of the angular PS and sky map, which, due to the very limited number of function evaluations, is much more computationally efficient than methods based on map generation and inversion or based on the standard $a_{\ell,0} (\nu,\beta)$ calculation which inverts the spherical harmonic expansion via integrals. This greatly alleviates the computation efforts needed in many applications requiring repeated calculations for different models and/or observer velocities.
In particular, Eq. \eqref{eq:struct_evenodd} can be directly used for background monopole spectra that are accurately described by functions that are analytical, semi-analytical or that, although tabulated, are numerically extremely accurate. The situation is different, in both this approach and in other ones, in many cosmological and astrophysical contexts in which these functions are affected by non-negligible numerical uncertainties.
This is due -- in Eq. \eqref{eq:struct_evenodd} -- to the potential imbalance between the slightly different $T_{\rm th}^{\rm BB/dist} (w)$ signals at different $w$ or -- in Eq. \eqref{eq:struct_der_ell} -- to the well known increase of the numerical uncertainty impact with increasing derivative order. The proposed method can be extended by applying appropriate filtering methods to the function describing the background monopole spectrum and/or to its derivatives. Depending on the spectral shape (smooth or feature-rich), on the level of numerical uncertainties and on the multipole of interest, accurate predictions can be efficiently carried out\cite{2024A&A...684A..82T} applying ($a$) a pre-filtering to the background monopole spectrum via a low-pass Gaussian filter in Fourier space and/or ($b$) a low-pass Gaussian filter in real (or Fourier) space, in sequence, to the monopole spectrum derivatives and/or ($c$) a dedicated approach that performs the $a_{\ell,0} (\nu,\beta)$'s calculation with amplified values $\beta_a$ and then scales the result to the true $\beta$ value by using the above relations.\footnote{The extremely accurate computation of $R_n$ would require to accurately know the change of the $n$-order derivative in the extremely narrow range ${\nu'_{\beta,\perp}} \div {\nu'_{\beta_a,\perp}}$, which, because of numerical uncertainties, is obviously missing, calling for filtering methods. In practice, one can set $R_n$ = 1.}

\subsection{Observational perspectives}
\label{sect:spectrumobs}

The observations of the background monopole spectrum are typically performed through dedicated absolute measurements, that clearly take advantage from the amplitude of the full global signal. On the other hand, as discussed above, the observer peculiar motion offers the chance to investigate the monopole spectrum also through the frequency dependence of the signal variation in the sky. This approach has been deeply investigated in the context of future nearly all-sky CMB differential experiments as a way to circumvent difficulties in absolute calibration and to deal with the foreground impact that, in particular, could be less relevant in dipole than in monopole analyses. The impact in dipole analyses of potential residuals from imperfect calibration and foreground subtraction has been addressed in Ref. \citenum{2018JCAP...04..021B} while a conceptual design of instrument and measurement techniques for the final output calibration has been discussed in Ref. \citenum{2018MNRAS.477.4473M}. 
The expected improvement with respect to COBE/FIRAS in the recovery of CMB distortion parameters\cite{1996ApJ...473..576F} and in the amplitude of the CIB spectrum\cite{1998ApJ...508..123F} could in principle be $\approx 10^3$ for an ideal differential experiment with a Cosmic Origins Explorer (CORE) like configuration,\cite{2018JCAP...04..014D} collecting a few thousands of receivers at the focal plane of a $\sim 1.2$m class telescope.\footnote{Even an experiment like the Lite (Light) Satellite for the studies of B-mode polarization and Inflation from cosmic background Radiation Detection\cite{2016SPIE.9904E..0XI} (LiteBIRD), although optimized for CMB polarization, could improve current constraints, depending on the quality of its total intensity anisotropy measurements.} Even in the conservative case of only 1\% accuracy (at a reference scale of about $1^\circ$) in both foreground removal and relative calibration, such a differential project would be able to improve the recovery of the CIB spectrum amplitude of a factor of $\simeq 17$, to achieve a marginal detection of the energy release associated with astrophysical reionisation, which produces a global Comptonisation distortion,\cite{1972JETP...35..643Z} and to improve by a factor of $\simeq 4$ the limits on early energy dissipations, resulting in Bose-Einstein like distortions,\cite{1970Ap&SS...7...20S} possibly associated to the damping of primordial perturbations. More refined foreground mitigation and calibration would greatly improve our knowledge of CIB spectrum and of the constraints -- or the detection accuracy -- of the $u$ Comptonization parameter and of the Bose-Einstein chemical potential, $\mu$, ranging from the conservative to the ideal case.\cite{2018JCAP...04..021B}

\begin{figure}[t!]
\centering
    \begin{minipage}[t]{0.49\linewidth}
        \includegraphics[width=5.5cm]{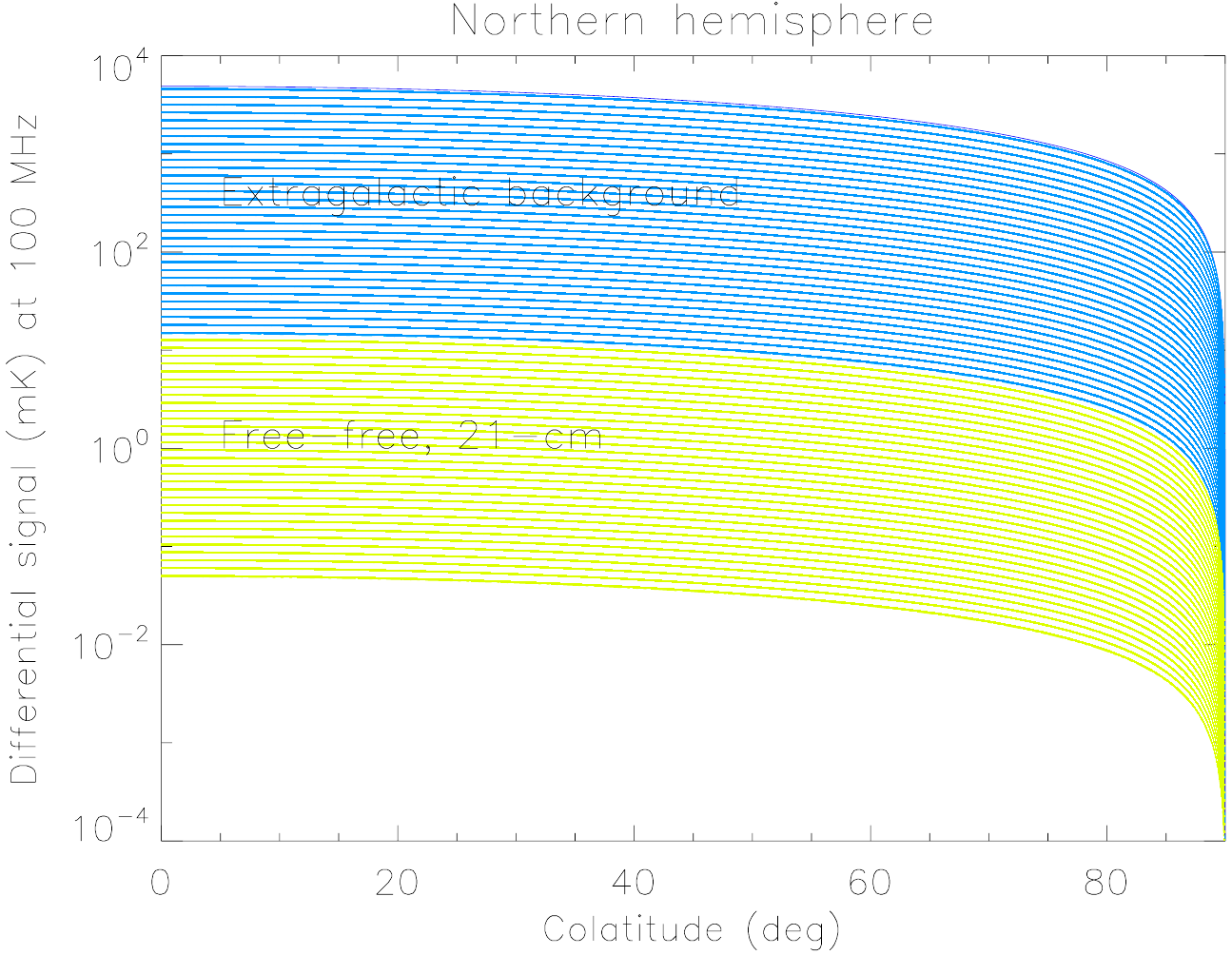}
    \end{minipage}
    \begin{minipage}[t]{0.49\linewidth}
        \includegraphics[width=5.6cm]{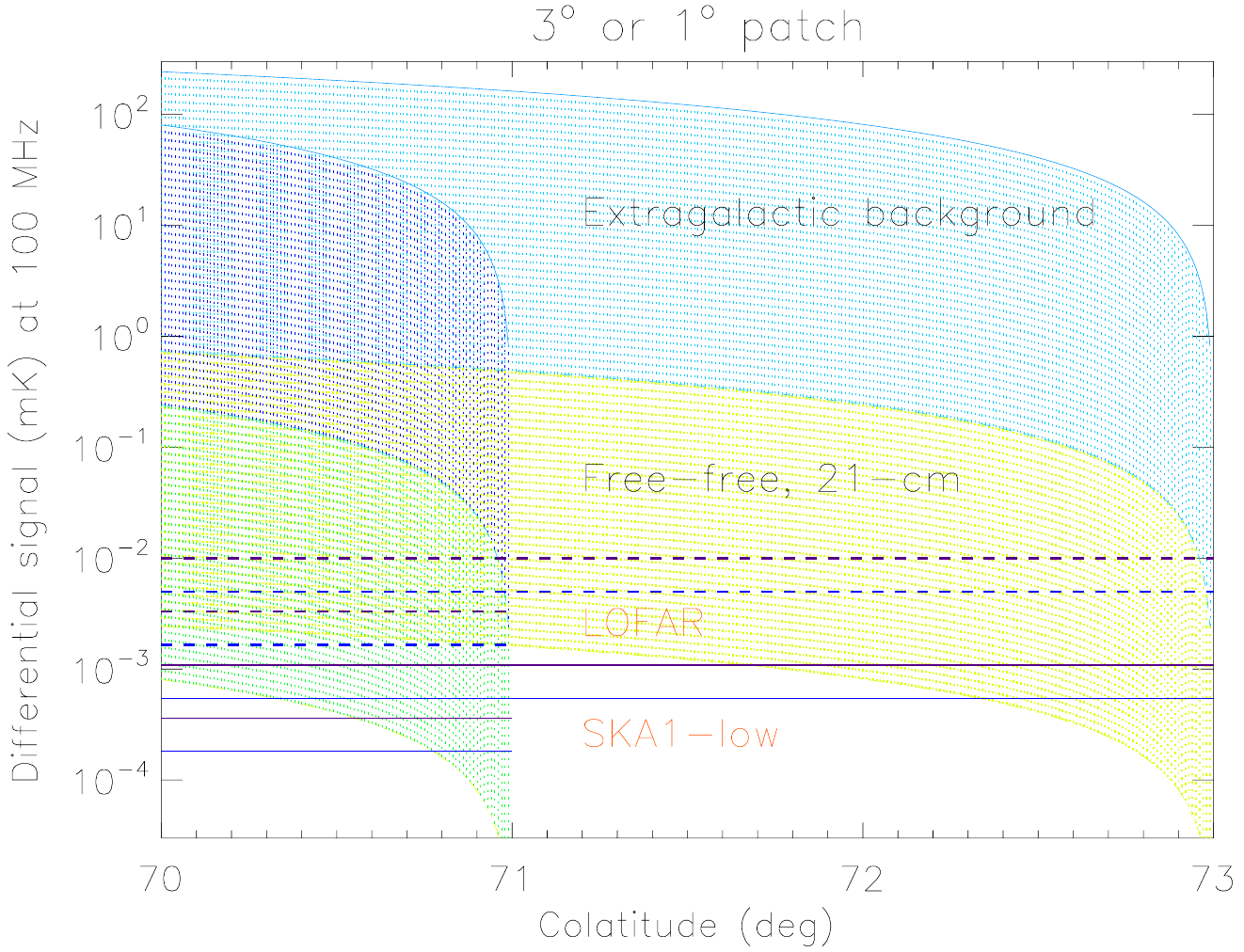}
    \end{minipage}  
    \caption{Range of predicted differential signal for the dipole, $\Delta T(\theta) = \Delta a_{\ell,0} $ $[3/(4\pi)]^{1/2} $ $ {\rm cos} \theta$, where $\theta$ is the colatitude and $\Delta a_{\ell,0}$ refers to the dipole harmonic component $\ell$\,=\,1, $m$\,=\,0 in a frame with the $z$-axis parallel to the observer motion, after the subtraction of the standard CMB BB, in order to emphasize the interesting signal (here at 100 MHz and in equivalent thermodynamic temperature). Left panel: the case of an all-sky survey, displayed for simplicity only for a hemisphere. Right panel: a zoom of left panel for a patch of 3$^{\circ}$ (1$^{\circ}$); here we display $\Delta T(\theta) - \Delta T(\theta_{*})$, with $\theta_{*}$ = 73 $^{\circ}$ (71$^{\circ}$), i.e. the differential signal inside the patch, to be compared with typical sensitivity levels\cite{dewdney2016} for LOFAR (dashed lines) and SKA1-low (solid lines) in a nominal pixel of 2 arcmin for one day of integration in the patch. Violet (blue) lines assumes a bandwidth of 10 (40) MHz to appreciate spectral shapes of the 21cm redshifted HI line (the other types of signal).
    }
    \label{fig:dippatch}
\end{figure}

For the $z$-axis parallel to the observer velocity, the amplitude of the observable background signal variation depends on the range of colatitudes considered. This is shown for the dipole in Fig. \ref{fig:dippatch}:
while nearly all-sky differential observations are sensitive to the whole dipole amplitude, the signal variation appears also at small scales essentially as a signal gradient along the meridian. This implies that for experiments with extremely high sensitivity and resolution the signal variation can be significant in comparison to the experiment sensitivity even considering a small sky patch:\cite{2019A&A...631A..61T} this is the case, in particular, of on-going and forthcoming interferometric projects at radio wavelengths. For example, Fig. \ref{fig:dippatch} compares the sensitivity of LOFAR and SKA1-low at 100\,MHz with the typical range of the differential dipole signal on a patch of
3$^{\circ}$ or 1$^{\circ}$ side for three types of cosmic background signal relevant at radio frequencies: the integrated background produced by extragalactic sources (see Sect. \ref{sect:ES}), the cosmological diffuse FF emission and the redshifted 21cm line global signal from HI (see Sect. \ref{sect:EoR}). For the extragalactic source background the signal range mainly reflects different levels of source detection threshold, for the 21cm line the range is related to the large envelope of the current model predictions,\cite{pritchard_loeb_2012,2017MNRAS.472.1915C} while for the FF it comes from considering minimal models accounting for the diffuse emission from the IGM during reionisation\cite{2014MNRAS.437.2507T} or the integrated emission from ionised halos.\cite{1999ApJ...527...16O} Of course, collecting many patches would improve the signal to noise ratio.
As evident, LOFAR and SKA1-low in principle have the chance to detect or even characterise these types of dipole signal (see Ref. \citenum{Trombetti_thisproc} for further details).

Future work will focus on selecting the favourite locations of an ensemble of sky patches, understanding the impact of Galactic foregrounds and the required quality of their mitigation, and defining suitable strategies to analyse the data and their calibration keeping the relevant (dipole and beyond) information while controlling systematic effects.
Together with dedicated absolute measurements, this type of analysis can help understanding controversial questions about the CRB spectrum (see Sect. \ref{sect:back}) and, in combination with other types of dipole analyses, to perform new tests about the geometrical properties of the Universe.

\section{Ultra-bright high-redshift strongly lensed galaxies}
\label{sect:lens}

One of the most notable discoveries from the {\it Planck} surveys was the unexpected -- though previously predicted by Ref. \citenum{Negrello2007} -- detection of ultra-bright, strongly lensed high-$z$ sub-mm galaxies with extreme magnifications ($\mu \sim 10\div50$; see Refs. \citenum{Canameras2015} and \citenum{Harrington2016}). Although {\it Planck} lacks the angular resolution to resolve their detailed structures, it can detect them, and then follow-up observations with higher resolution facilities can shed light on their nature and on their properties. These observations can offer unique insights into the internal structure and kinematics of high-$z$ galaxies during their most active star formation phases, shrouded in dust. Studying these sources is key to understanding the fundamental processes driving early galaxy formation and evolution. Current models suggest various mechanisms, including mergers, interactions, cold gas flows, and in situ processes (see Refs. \citenum{Silk2012} and \citenum{Somerville2015}), but rely on many adjustable parameters to fit statistical data such as source counts and redshift distributions (see Sect. \ref{sect:ES}).

Probing the interiors of high-$z$ SFG is crucial to directly observe these processes. However, these galaxies are typically compact, with sizes of $1\div2$ kpc (corresponding to\cite{Fujimoto2018} $0.1\div0.2$ arcseconds at $z \simeq 2\div3$), which makes them difficult to resolve even with high-resolution instruments like ALMA, HST and JWST. Adequate resolution and signal-to-noise ratios are generally achieved only for the brightest galaxies, which may not represent the general population.

Gravitational lensing provides a solution, enabling the study of high-$z$ galaxies in greater detail than current instruments allow. It magnifies both the flux and the apparent size of the galaxy, stretching it by an average factor of $\mu^{1/2}$, while conserving the surface brightness. In the case of extreme magnifications, ALMA's $0.1''$ resolution, for instance, allows spatial resolutions reaching several tens of parsecs, i.e. sizes comparable or smaller than those of Galactic giant molecular clouds. Spectroscopy, like that from Ref.\citenum{Canameras2017}, has achieved uncertainties of $40–50\, \rm km/s$, enabling the accurate study of high-$z$ AGN driven outflows, having velocities of $\sim 1000\, \rm km/s$.\cite{King2015}

Observing and measuring outflows is essential for galaxy formation theory, trying to explain why only 10\% of baryons form stars. However, observational evidence at high-$z$ is limited. Strong lensing allowed Ref. \citenum{Spilker2018} to detect a fast ($800\, \rm km/s$) molecular outflow in a galaxy at $z = 5.3$.

\subsection{Using Planck maps for a systematic search}
\label{sect:lenswithPlanck}

A statistically significant sample across a broad redshift range is needed to draw reliable conclusions about galaxy formation. So far, {\it Planck} has detected a few dozen lensed galaxies, mostly through the Herschel Director's Discretionary Time (DDT) ``Must-Do'' Programme\cite{Canameras2015} and cross-matching {\it Planck} catalogues with large-area sub-mm surveys.\cite{Berman2022} Given the limited sky coverage, many such objects remain undiscovered. 
Although {\it Planck} cannot resolve arcs or Einstein rings, it can detect strongly lensed galaxies through their sub-mm emission. Most strongly lensed galaxies are detected at frequencies $\nu \geq 353$\,GHz,
with {\it Planck} sub-mm surveys identifying galaxies at $z > 1$, showing colder sub-mm colors than local galaxies.\cite{Negrello2017}

\begin{figure}[t!]
\begin{center}
\includegraphics[width=11cm]{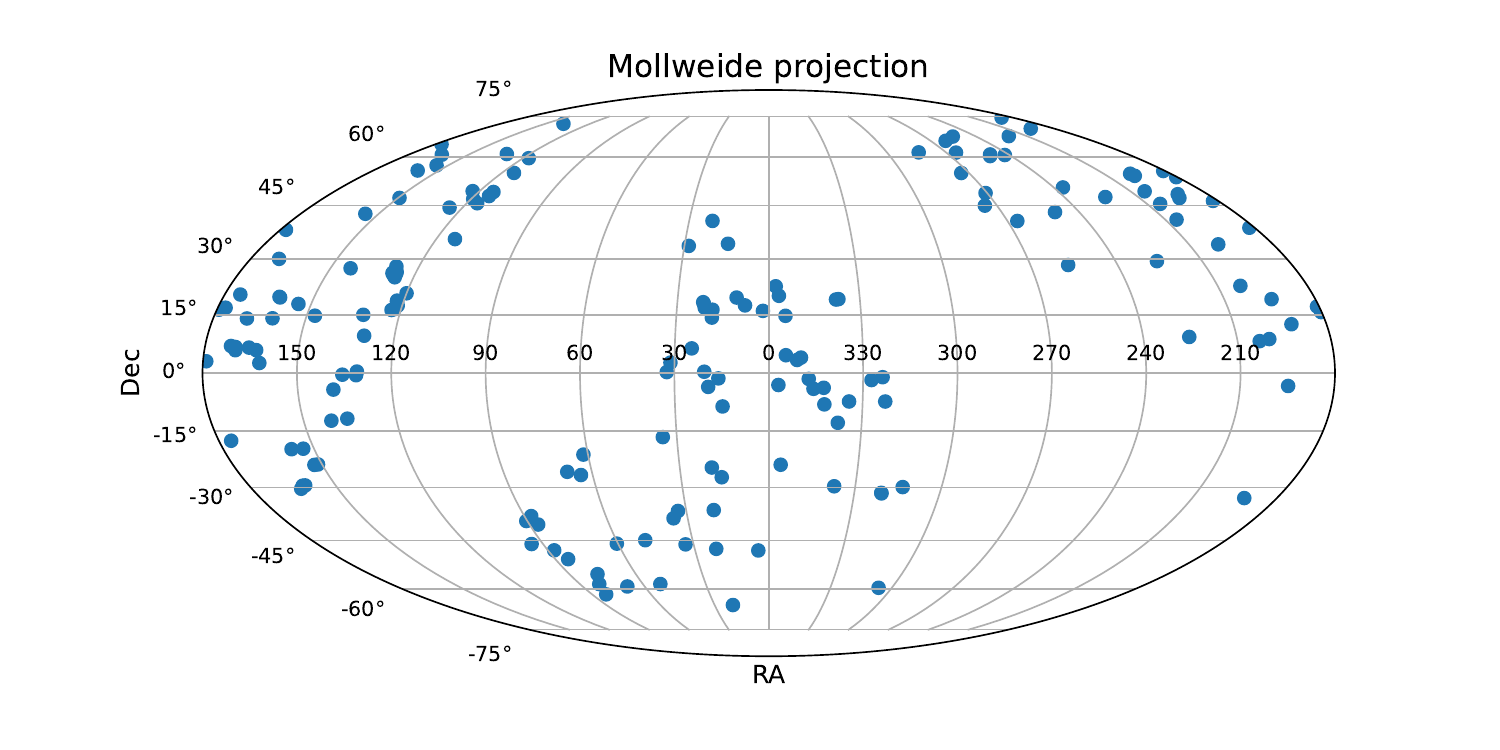}
\caption{Mollweide projection of the spatial distribution of candidate strongly lensed galaxies selected from the {\it Planck} 353\,GHz catalogue.}
\label{fig:Mollweide}
\end{center}
\end{figure}

A simple estimate\cite{Trombetti2021} suggests that {\it Planck} catalogues may contain around 150 strongly lensed galaxies at high Galactic latitude ($|b|> 20^\circ$). These would constitute an excellent sample for studying galaxy structure across a broad redshift range. The sample selection process for lensed candidates from {\it Planck} maps is detailed in Ref. \citenum{Trombetti2021}. Briefly, the Second {\it Planck} Catalogue of Compact Sources (PCCS2) at the three highest frequencies (353, 545, and 857 GHz) is used, focusing on sources with Galactic latitudes $|b|>20\,\rm deg$ to minimize Galactic contamination. Nearby galaxies are filtered by cross-matching with 
InfraRed Astronomical Satellite (IRAS) catalogues and radio sources are excluded using the {\it Planck} multi-frequency catalogue of non-thermal sources (PCNT).\cite{PlanckCollaborationPCNT} A further refinement was done by sub-mm color analysis and multi-frequency signal enhancement using the Matrix Filters methodology. 104, 177, and 97 candidates have been identified at 353, 545, and 857 GHz, respectively. The spatial distribution of the sample selected at 353\,GHz is shown, as an example, in Fig.\,\ref{fig:Mollweide}.

Machine-learning techniques could further refine this approach by generating a tailored training set built using the {\it Planck} Sky Model simulations.

It is probable that over 50\% of the sources in the sample are not strongly lensed galaxies but instead a mix of other objects with cold SEDs.
These include Galactic cold clumps, Galactic cirrus, high-$z$ proto-clusters of dusty galaxies, and fluctuations in the CIB.

Identifying strongly lensed galaxies using {\it Planck} data alone is almost hopeless, particularly due to its low angular resolution, which leads to significant positional uncertainties, of one arcminute or more for these low signal-to-noise sources. To address this, follow-up observations using the Australia Telescope Compact Array (ATCA), 
the second generation Neel-IRAM-KID-Array (NIKA 2) and the Submillimetre Common-User Bolometer Array 2 (SCUBA-2) have been conducted. In addition, four of the detected sources underwent spectroscopic observations with the NOrthern Extended Millimeter Array (NOEMA). Every object in the sample has been observed by at least one of these instruments. These follow-up observations confirm the presence of the sources and provide precise positional information, enabling future high-resolution studies using telescopes such as ALMA, JWST and the Karl G. Jansky Very Large Array (JVLA).

The analysis of these follow-up observations is in progress, with some preliminary findings by Trobbiani et al. (in prep.) suggesting the successful identification of new strongly lensed galaxies. These discoveries will ultimately contribute to a deeper understanding of the internal structure, star formation rates, and kinematics of high-$z$ galaxies.

While designed for cosmology, {\it Planck} has proven valuable in finding ultra-bright high-$z$ lensed galaxies. However, detecting lensed galaxies with {\it Planck} is challenging due to low signal-to-noise ratios, inaccurate positions and blurred photometric properties. The proposed systematic approach (see Ref. \citenum{Bonato_etal_thisproc} for further details), combined with follow-up observations, has the potential to significantly expand the known sample, offering critical insights into galaxy formation and evolution during the peak of cosmic star formation. The method outlined has the potential to increase the number of detected galaxies by a factor $\simeq 3\div4$, though with limited efficiency.

Future work will focus on refining the selection process, exploring machine-learning alternatives, and conducting in-depth studies of confirmed lensed galaxies.

{\small
\section*{Acknowledgments}

MB, CB and TT acknowledge support from INAF under the mini-grant ``A systematic search for ultra-bright high-z strongly lensed galaxies in {\it Planck} catalogues''.
CB acknowledges support from the InDark INFN Initiative (\url{https://web.infn.it/CSN4/IS/Linea5/InDark/index.html}). LT acknowledges the Spanish Ministerio de Ciencia, Innovaci\'on y Universidades for partial financial support under the projects PID2022-140670NA-I00 and PID2021-125630NB-I00.}


\end{document}